# Large Language Model Powered Decision Support for a Metal Additive Manufacturing Knowledge Graph

**Muhammad Tayyab Khan[1,3*], Lequn Chen[2], Wenhe Feng[1] and Seung Ki Moon[3]**

[1] Singapore Institute of Manufacturing Technology (SIMTech), Agency for Science, Technology and Research (A*STAR), 5 CleanTech Loop, #01-01 CleanTech Two Block B, Singapore 636732, Republic of Singapore

[2] Advanced Remanufacturing and Technology Centre (ARTC), Agency for Science, Technology and Research (A*STAR), 3 CleanTech Loop, #01-01 CleanTech Two, Singapore 637143, Republic of Singapore

[3] School of Mechanical and Aerospace Engineering, Nanyang Technological University, 639798, Singapore

* Corresponding Author / Email: KHAN0022@e.ntu.edu.sg

*Metal additive manufacturing (AM) involves complex interdependencies among processes, materials, feedstock, and post-processing steps. However, the underlying relationships and domain knowledge remain fragmented across literature and static databases that often require expert-level queries, limiting their applicability in design and planning. To address these limitations, we develop a novel and structured knowledge graph (KG), representing 53 distinct metals and alloys across seven material categories, nine AM processes, four feedstock types, and corresponding post-processing requirements. A large language model (LLM) interface, guided by a few-shot prompting strategy, enables natural language querying without the need for formal query syntax. The system supports a range of tasks, including compatibility evaluation, constraint-based filtering, and design for AM (DfAM) guidance. User queries in natural language are normalized, translated into Cypher, and executed on the KG, with results returned in a structured format. This work introduces the first interactive system that connects a domain-specific metal AM KG with an LLM interface, delivering accessible and explainable decision support for engineers and promoting human-centered tools in manufacturing knowledge systems.*

## 1. Introduction

Metal additive manufacturing (AM) is a knowledge rich domain characterized by complex interdependencies among materials, processes, machine configurations, and post-processing operations [1]. Optimal decision-making in this context requires not only identifying material and process compatibility but also accounting for feedstock constraints, thermal histories, and downstream treatment requirements. Capturing such multifaceted information in a machine-readable format can significantly enhance manufacturability analysis, design iteration, and process planning.

Knowledge graphs (KGs) have emerged as a powerful paradigm for modeling structured engineering knowledge. By semantically linking entities such as materials, AM processes, and post-processing techniques, KGs preserve contextual and relational information more effectively than traditional databases and support more expressive querying. However, while the utility of KGs in smart manufacturing and digital twins is increasingly recognized, their application in the specific context of metal AM remains limited and disconnected from real-time decision-making workflows [2].

Existing AM knowledge systems, including ontology based models [3], [4], [5] and industrial repositories such as Senvol Database [6], provide valuable data but are constrained by static access mechanisms. These systems often require users to master formal query languages or navigate complex interfaces, limiting their usability during rapid design and production cycles. Moreover, they lack natural language interfaces, thereby excluding non-expert users from effectively interacting with domain knowledge.

An additional limitation is the lack of real-time reasoning and the ability to evaluate across multiple constraints in current AM knowledge platforms. Most existing systems are designed for passive information retrieval rather than interactive decision support [2]. As a result, design engineers rely on fragmented resources or domain experts to interpret data, which reduces productivity and limits knowledge reuse.

To address these challenges, we propose a novel decision support framework that integrates a manually curated and semantically rich metal AM KG with a few-shot prompted large language model (LLM) to enable natural language querying. Unlike prior approaches, our system allows users to pose flexible conversational queries such as process selection, feedstock filtering, and design for AM (DfAM) constraints, without requiring knowledge of Cypher or SPARQL. The LLM is guided by schema-aware and domain-specific prompts that translate user queries into Cypher, which are then executed over a Neo4j based KG containing 53 distinct metals and alloys, nine AM processes, four feedstocks, and post-processing attributes [7]. This work presents, to the best of our knowledge, the first real-time interactive system that connects a domain-specific metal AM KG with an LLM interface, enabling intuitive, explainable, and accessible decision support for engineering applications.

## 2. Methodology

### 2.1 Knowledge Graph Construction for Metal AM

To support structured and context-aware decision-making in metal AM, we construct a novel, domain-specific KG in Neo4j using Cypher language. The KG semantically encodes interrelated information across materials, AM processes, feedstock types, fusion techniques, and post-processing operations. Each node represents a physical or conceptual entity, while edges define compatibility, requirements, or transformation relationships critical to manufacturability.

A representative subgraph is shown in Fig. 1, illustrating the pathway from a sample material (Ti-6Al-4V) through compatible AM processes, feedstock formats, fusion techniques, and material state transitions. This layered structure captures not only printability and feedstock to process mappings but also dependencies such as thermal or mechanical treatment requirements after fabrication.

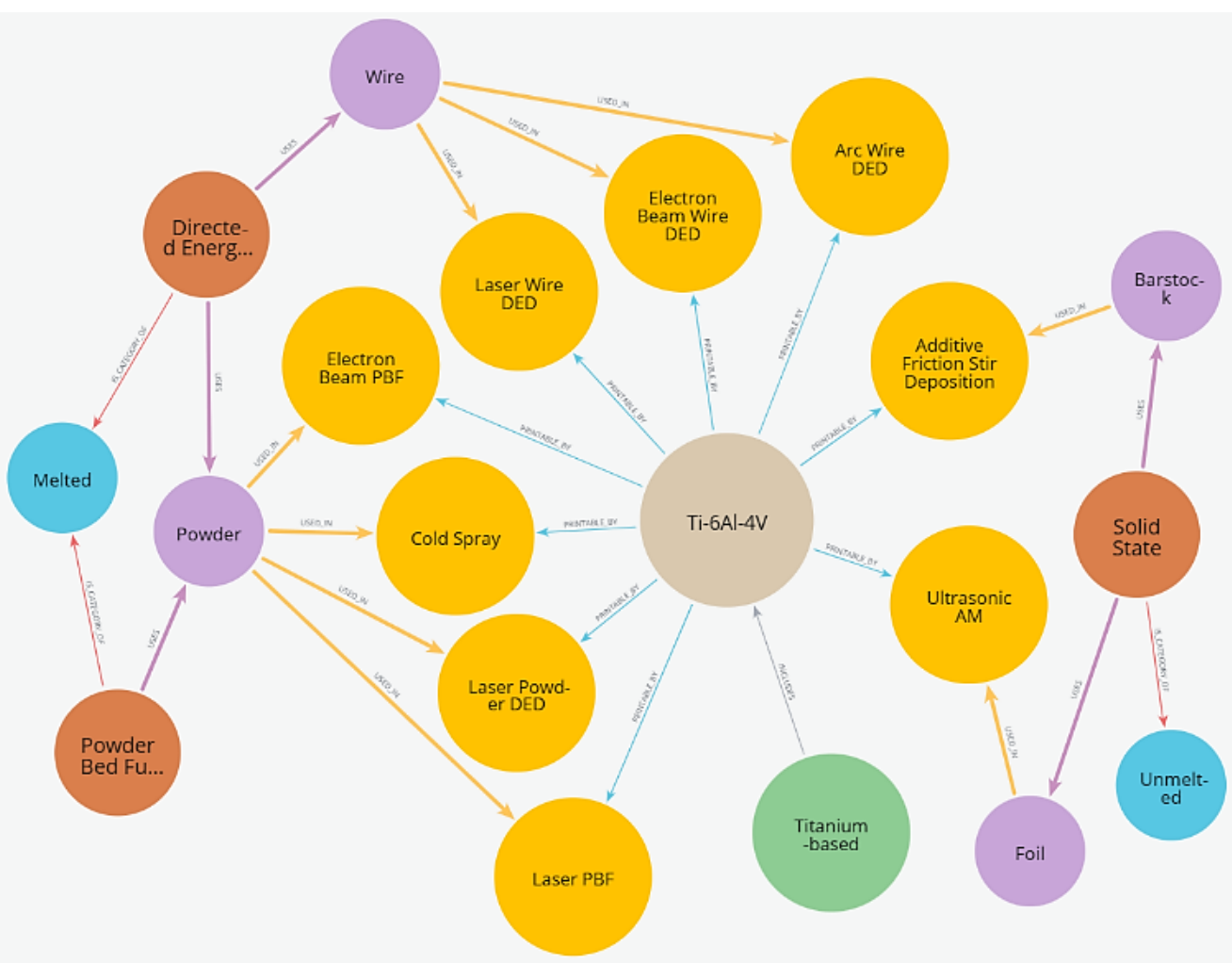

**Fig. 1.** *A representative subgraph from the metal AM KG, showing the interconnected relationships for Ti-6Al-4V. The figure highlights compatible AM processes, feedstock forms, fusion techniques, and material state transitions, reflecting the KG's layered structure.*

The complete KG includes:

- 53 distinct metals and alloys, grouped into seven families: nickel, iron, copper, cobalt, titanium, aluminum-based alloys, and refractory materials
- Nine AM process categories, including Laser Powder Bed Fusion (PBF), Electron Beam PBF, Directed Energy Deposition (DED) methods (laser, wire, arc), Cold Spray, Additive Friction Stir Deposition, and Ultrasonic AM
- Four feedstock formats: powder, wire, foil, and bar stock
- Post-processing steps, such as stress relief, powder removal, heat treatment, support removal, and final machining

Material to process compatibility is derived from verified industry mappings. Process nodes include detailed quantitative attributes such as deposition rate, feature resolution, and maximum build volume. These relationships allow the KG to support advanced queries involving multi-constraint filtering, such as identifying printable materials for a target geometry or selecting processes with specific post-processing needs.

Compared to existing AM ontologies and data repositories, our KG is the first to unify such detailed multi-domain knowledge into a single semantic model that is directly executable for reasoning over process, material, and feedstock interactions in metal AM.

### 2.2 Natural Language Querying with an LLM and Few-Shot Prompting

To make the structured KG accessible through intuitive user inputs, we integrate an LLM driven natural language interface. Rather than requiring Cypher syntax, users pose design and manufacturing queries in text, which are interpreted by a lightweight few-shot prompting framework using OpenAI's GPT-4o-mini.

Few-shot prompt offers a scalable and low maintenance alternative to rule-based templates and model fine-tuning. Each prompt contains a concise schema overview, nodes and relationships definitions, and synonym mappings to support accurate query interpretation. It is supplemented by more than 50 curated query response examples across eight functional categories that span both retrieval and reasoning tasks:

- **Basic Retrieval**: Listing entities such as AM processes, feedstock types, or material categories
- **Printability Analysis**: Determining which materials can be printed using a specific AM process
- **DFAM Guidance**: Retrieving process specific constraints such as build dimensions or feature resolution
- **Feedstock Engineering:** Exploring feedstock formats and size requirements across processes
- **Post-processing Estimation:** Estimating requirements such as heat treatment, stress relief, or support removal
- **Cross Material Compatibility:** Identifying processes capable of handling multiple material families
- **Capability Filtering:** Selecting processes based on thresholds such as build height, resolution, or deposition rate
- **Analytical Queries:** Performing multi-constraint reasoning to recommend suitable material and process pairs

At inference time, user queries are first normalized using a preprocessing module based on spaCy, which expands synonyms and resolves entities. The processed input is injected into the few-shot prompt, guiding the LLM to generate a corresponding Cypher query. This query is executed on the Neo4j graph, and the output is reformatted into a structured, user readable response using standardized units and domain consistent terminology.

To minimize hallucinations and enhance robustness, the prompt also includes negative examples, guiding the model to respond with "unsupported query" when user intent exceeds the current scope of the KG (e.g., queries involving unencoded mechanical properties). Since the graph currently contains hundreds of structured triples, all prompting components remain within token limits, ensuring full context availability during generation. The prompt also functions as both implicit documentation and a modular framework extensible to other AM domains. Fig. 2 outlines the complete pipeline, from natural language input, through query normalization and prompt-driven translation, to Cypher execution and structured response generation via a Streamlit-based chatbot interface.

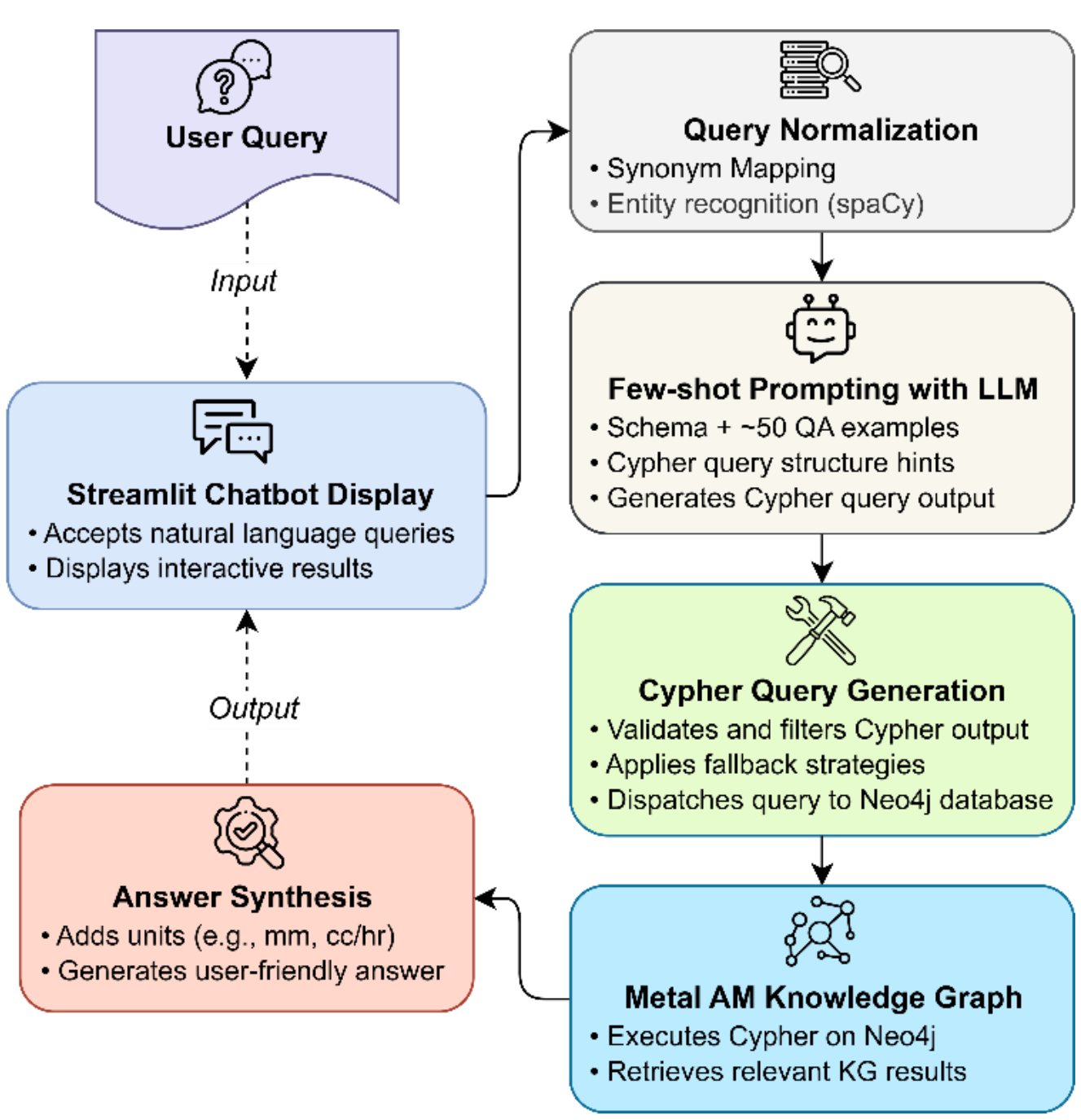

**Fig. 2.** *Architecture of the LLM based querying framework for the metal AM KG. User queries are normalized and processed by a few-shot prompting engine using GPT-4o-mini, which generates Cypher queries executed on Neo4j. Results are post-processed into structured, natural language outputs and delivered via a Streamlit chatbot interface.*

## 3. Results and Discussions

To evaluate the effectiveness of the proposed natural language querying system for the metal AM knowledge base, we developed a series of test cases reflecting realistic design and manufacturing scenarios. These cases were designed to assess the system's ability to interpret diverse query structures, perform multiple constraint filtering, and respond appropriately when relevant information was absent from the KG. The deployed interface allows users to submit queries through a Streamlit based chatbot, where inputs are first normalized and then transformed into Cypher queries using GPT-4o-mini, guided by the few-shot prompting framework. The generated queries are executed on the Neo4j based KG, and the results are reformatted into natural language responses.

Fig. 3 presents a successful material compatibility example. When prompted to identify alloys printability using a specific AM process, the chatbot returned a complete and accurate list based on the PRINTABLE_BY relationships encoded in the KG. The output format adheres to the structure established in the prompt, demonstrating the system's ability to extract multiple relevant entities and present grounded, non-hallucinated responses.

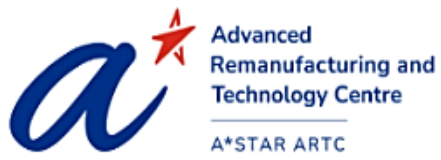

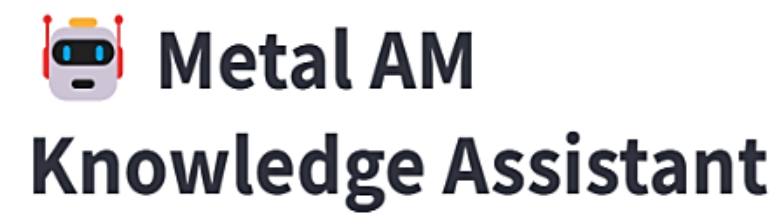

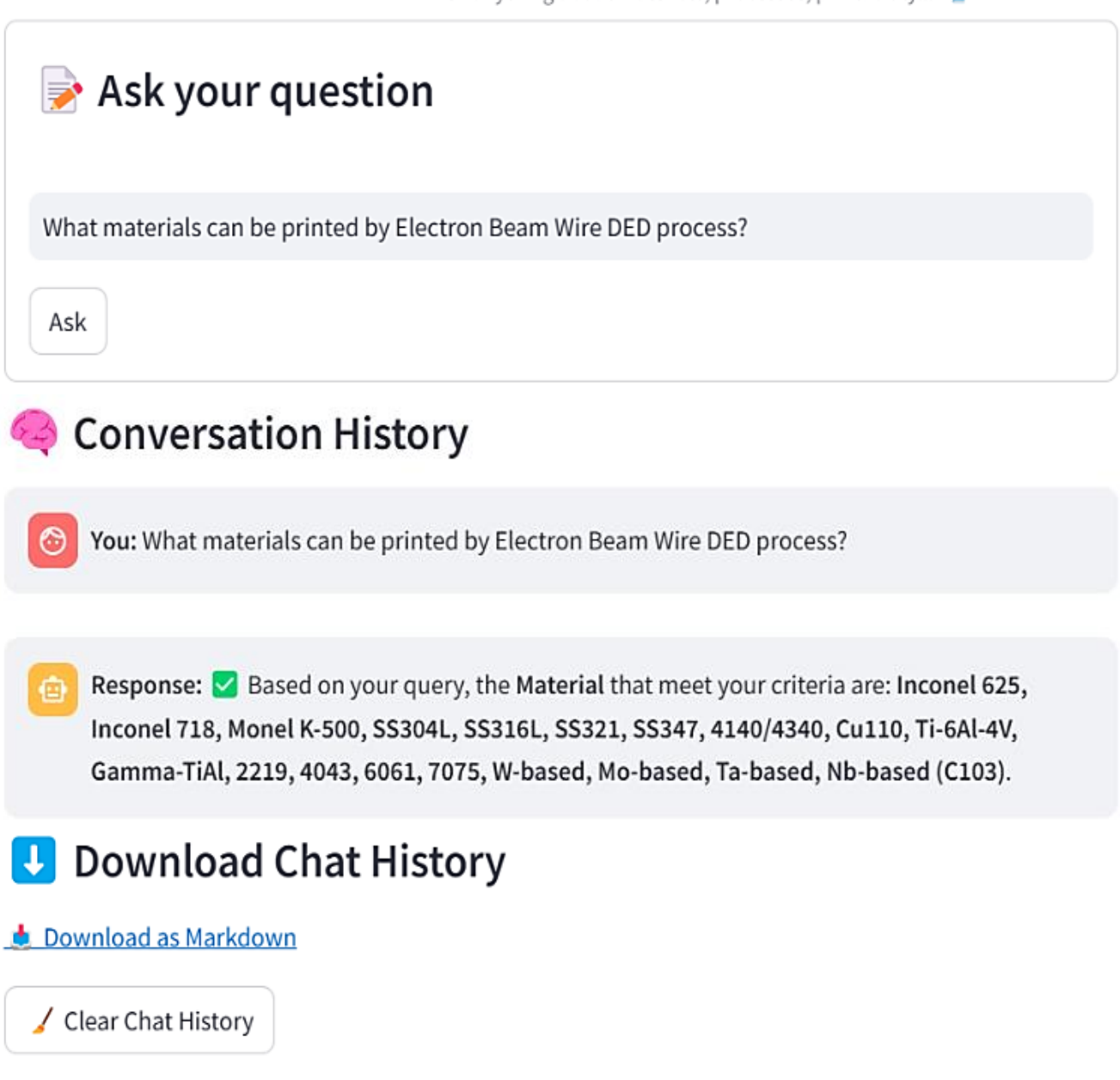

**Fig. 3.** *Chatbot response to a material compatibility query. The system correctly identifies all alloys printable by the Electron Beam Wire DED.*

Another use case is shown in Fig. 4, where the query involves identifying AM processes compatible with a specific alloy and requiring two simultaneous post-processing steps. The system successfully returned the correct process set, illustrating its ability to interpret compound constraints and perform logical filtering over graph relationships. The response was both complete and clearly structured, aligned with the expected prompt-driven output.

**Q:** Find metal AM processes needed for printing nickel-based alloys and requires both powder removal and heat treatment?

**A:** ✅ Based on your query, the metal AM processes that meet your criteria are: Laser Powder DED, Laser PBF, Electron Beam PBF.

**Fig. 4.** *Chatbot response to a compound filtering query. The system identifies AM processes compatible with nickel-based alloys that require both powder removal and heat treatment.*

To evaluate system boundaries, we submitted a deliberately unsupported query requesting mechanical behavior data across build orientations and post-processing treatments. As shown in Fig. 5, the system returned an appropriate rejection message, correctly identifying that such information is not represented in the current KG. This result underscores the effectiveness of embedding negative examples into the prompt, enabling the model to abstain from speculative or unsupported answers.

**Q:** Which AM processes exhibit anisotropic mechanical behavior in tensile strength across build orientations for Inconel 718 and Ti-6Al-4V under varying post-processing heat treatments?

**A:** ❌ Sorry, the current knowledge graph does not support this type of query.

**Fig. 5.** *An unsupported query: the system appropriately declines to answer due to the absence of mechanical behavior data in the KG.*

These results indicate that the system can successfully address a wide range of DfAM queries. In most cases, responses are accurate, contextually relevant, and grounded in the underlying knowledge base. Partial errors were occasionally observed in scenarios involving layered reasoning or when user phrasing diverged significantly from prompt examples. In such cases, the model might overlook part of the query or return incomplete results.

The overall performance of the system is primarily determined by two factors: the scope and level of detail represented within the KG, and the linguistic diversity and representativeness of the prompting examples. When these elements align with the user's intent, the model delivers precise and actionable answers. Conversely, gaps in schema coverage or misalignment in phrasing can degrade response quality. Despite these limitations, the LLM based interface offers an accessible and low barrier mechanism for querying structured metal AM knowledge, eliminating the need for formal query languages and reducing reliance on domain experts during early-stage design and planning.

## 4. Conclusions

This work presents a novel decision support framework that integrates a domain-specific and semantically rich metal AM KG with a natural language interface powered by a prompted LLM. The system enables intuitive access to complex and interrelated AM knowledge without requiring formal query language expertise. By leveraging carefully structured prompts and schema aware query transformations, the system provides accurate responses to diverse queries related to material and process compatibility, feedstock engineering, DfAM constraints, and post-processing planning.

The main contributions of this work are: (i) the construction of a comprehensive KG covering 53 alloys, nine AM processes, four feedstock types, and key post-processing operations; (ii) the design of a lightweight few-shot prompting strategy for natural language translation without LLM fine-tuning; (iii) the development of a real-time querying pipeline, deployed via a Streamlit interface; and (iv) a qualitative evaluation demonstrating robust performance across compatibility checks, multi-constraint reasoning, and unsupported queries.

This approach lowers the barrier to expert-level AM knowledge and provides accessible decision support for engineers and students in engaged in early-stage design and planning. While the system proves effective across a range of query types, its performance is shaped by the coverage of the KG and the scope of prompt examples. As the graph evolves to incorporate more detailed information, such as mechanical properties, design rules, or uncertainty representations, future work will explore LLM fine-tuning, multi-turn dialogue capabilities, and integration with CAD and simulation environments. Overall, this framework offers a modular, scalable, and human-centered interface for intelligent decision support in metal AM, advancing the use of conversational AI in engineering knowledge systems and smart manufacturing.

## ACKNOWLEDGEMENT

This work is supported by the Agency for Science, Technology and Research (A*STAR), Singapore through the RIE2025 MTC IAF-PP grant (Grant No. M22K5a0045). It is also supported by the Singapore International Graduate Award (SINGA) (Awardee: Muhammad Tayyab Khan), funded by A*STAR and Nanyang Technological University, Singapore.